\documentclass[a4paper,12pt]{article}
\usepackage{amsmath}
\usepackage{amssymb}
\usepackage{amsthm}
\usepackage{algorithm}
\usepackage{algorithmic}
\usepackage{epsf}
\usepackage{epsfig}
\usepackage{graphicx,color,rotating}
\usepackage{multirow,longtable}
\usepackage{graphicx}
\usepackage{ucs}
\usepackage[applemac]{inputenc}
\usepackage{makeidx}  

\newtheorem{thm}{Theorem}[section]
\newtheorem{cor}[thm]{Corollary}
\newtheorem{lem}[thm]{Lemma}
\newtheorem{obs}[thm]{Observation}
\newtheorem{alg}[thm]{Algorithm}
\newtheorem{pro}[thm]{Procedure}

\theoremstyle{remark}
\newtheorem{rem}[thm]{Remark}

\theoremstyle{definition}

\begin{document}
\title {Minimizing the total weighted pairwise connection time in network construction problems }

\author{ Igor Averbakh$^1$\vspace*{0.25 cm} \\
$^1$ Department of Management, University of Toronto Scarborough\\
1265 Military Trail, Toronto, Ontario M1C 1A4, Canada\\
and Rotman School of Management, University of Toronto,\\
 105 St. George Street, Toronto, Ontario M5S 3E6, Canada\vspace*{0.25 cm}\\
email: igor.averbakh@rotman.utoronto.ca}

\maketitle

\textbf{Abstract}.   It is required to find an optimal order of constructing the edges of a network so as to minimize the sum of the weighted connection times of relevant pairs of vertices.   Construction can be performed anytime anywhere in the network, with a fixed overall construction speed. The problem is strongly NP-hard even on stars. We present polynomial algorithms for the problem on trees with a fixed number of leaves, and on general networks with a fixed number of relevant pairs. 


\textbf{Declaration of interests:} None.

\section{Introduction}\label{introduction} 

An increasing amount of literature is devoted to planning construction activities for building new transportation or communication networks, or restoring existing networks destroyed/damaged as a result of a disaster. Such activities are typically done with limited resources, hence scheduling issues arise: when and in what order to construct/restore different edges taking into account that some network connections may be more important than others and need to become functional before the whole network becomes operational. In a recently introduced class of network construction / connectivity restoration problems (\emph{NC problems}), the focus is on connectivity issues, and the goal is to find a schedule of construction activities that optimizes an objective which is a function of the times when certain pairs of vertices (relevant pairs) become connected. See \cite{averbakhJOC, averbakhCOR} for a review of related literature.

In this paper, we introduce and study a new NC problem. We consider the setting where due to limited resources the overall construction speed is fixed, and at any time construction can be conducted at any point(s) of the network. This is based on the assumption that the construction resources (crews, materials, equipment) can be transported from any point of the network to any other point, in time negligible with respect to construction times via some modes of transportation that do not use the network under construction. The objective is to minimize the sum of the weighted times when different pairs of vertices become connected (the total weighted connection time). The same setting but with a different objective (minimizing the maximum lateness of connection times with respect to a given set of due dates) was considered in \cite{averbakhJOC}. As we discuss in Section \ref{problem}, a problem which mathematically is equivalent to a special case of our problem was studied in \cite{averbakhpereira, alpern, hermans}.

Our focus in this paper is on computational complexity issues. First, we show that the problem is strongly NP-hard even on star networks (which are a very special case of trees), in contrast with the NC problems discussed in \cite{averbakhpereira, averbakhJOC, averbakhCOR} which are polynomially solvable on trees. Second, we develop a polynomial $O(n^{l+2})$ exact algorithm for the problem on trees with a fixed number $l$ of leaves ($n$ is the number of vertices). This, for example, translates into an $O(n^4)$ algorithm for the problem on a path. Third, for the problem on a general network with a fixed number $r$ of relevant pairs (vertex pairs with non-zero weights whose connection times are relevant for the objective function), we develop a polynomial $O(n^{2r-2})$ exact algorithm with an $O(n^3)$ pre-processing. For the important special case of all relevant pairs having a common vertex (which, as we show in Section \ref{problem}, is equivalent to the problem studied in \cite{averbakhpereira}, \cite{alpern} and \cite{hermans}), the complexity is reduced to $O(n^{r-1})$ with $O(n^3)$ pre-processing. The results for the problem with a fixed number of relevant pairs are applicable to other objectives that are monotonically non-decreasing functions of the connection times of the relevant pairs, e.g. to the NC problem studied in \cite{averbakhJOC}.

\section{The problem}\label{problem}

Suppose that a connected network $G=(V,E)$  with the set of vertices $V$, $|V|=n$, and the set of undirected edges $E$, $|E|=m$ needs to be constructed.   
Construction starts at time 0 and proceeds with a fixed construction speed; specifically,  1 unit of length of the network can be constructed per 1 unit of time.  At any time, construction activities can be performed at any points of the network, as long as the overall construction speed remains fixed.  (We assume that there are available modes of transportation  that do not depend on the network under construction, and that transportation times are negligible with respect to construction times, so construction crews, materials, and equipment at any time can be relocated from any point to any other point of the network instantaneously even if the two points are not connected by an already constructed path. A justification for this assumption is discussed in \cite{averbakhJOC}.)  The instant when two vertices $u,v$ become connected by an already constructed path is called the \emph{connection time} for $u,v$ and is denoted $t_{\{u,v\}}$. For any pair of vertices $\{u,v\}$, $u,v\in V$,  a nonnegative weight $w_{\{u,v\}}$ is given,  $w_{\{v,v\}}=0$ for any $v\in V$.   A vertex pair $\{u,v\}$ such that $w_{\{u,v\}}>0$ will be called a \emph{relevant pair}, or \emph{r-pair}, and $\cal R$ will denote the set of all r-pairs. Our purpose is to choose a construction schedule that minimizes the total weighted connection time $\sum_{\{u,v\}\in {\cal R}}w_{\{u,v\}}t_{\{u,v\}}$. 

A similar setting but with a different objective (minimizing the maximum lateness of vertex pairs' connection times with respect to some due dates) was considered in \cite{averbakhJOC}. It was observed in \cite{averbakhJOC} that since the overall construction speed is fixed, for any objective that is a monotonically nondecreasing function of the connection times, the following two properties hold:

a) ``Even if there is a possibility to perform construction at different places simultaneously, there is an optimal solution where at any time only one edge is being constructed. That is, it is not beneficial to split the limited resources between different edges." \cite{averbakhJOC}

b) ``There is an optimal solution without preemption, where construction of an edge is not interrupted once it has been started." \cite{averbakhJOC}

Therefore, we consider only schedules that satisfy properties a) and b) above. Such schedules consist of steps where at each step one edge is constructed. The edges that are constructed before all vertices become connected are called \emph{essential}. The order of constructing the remaining edges does not affect the objective value. It is clear that there is an optimal schedule where the essential edges form a spanning tree of $G$, and therefore are constructed in the first $n-1$ steps.  
A sequence of edges that form a spanning tree will be called an \emph{s-sequence} (solution sequence). Thus, the problem is to find an optimal s-sequence. This problem will be called \textbf{Problem A}.

Observe that for a spanning tree of essential edges \emph{any} order of constructing its edges is feasible.  
Thus, in an optimal schedule, at some instants the already constructed subnetwork may have more than one nontrivial connected components.

If there is only one r-pair $\{u,v\}$, then Problem A is equivalent to the problem of finding a shortest path between $u$ and $v$.
If all r-pairs have a common vertex $v$ (the \emph{depot}), then clearly there is an optimal solution where at any time the already constructed subnetwork is connected and contains $v$, hence this special case of Problem A is equivalent to the Flowtime Network Construction Problem (FNCP) considered by Averbakh and Pereira in \cite{averbakhpereira}. In FNCP, all construction resources are initially located at the depot and can be transported only within the already constructed part of the network, and it is required to minimize the sum of the weighted times when different vertices become connected to the depot. FNCP is strongly NP-hard on general networks \cite{averbakhpereira}, hence Problem A is strongly NP-hard as well. However, FNCP is polynomially solvable on trees 
\cite{averbakhpereira}; in contrast, as we show below, Problem A is strongly NP-hard even on so special case of trees as stars. There is an interesting connection between FNCP and the search theory: FNCP is equivalent to the \emph{expanding search} problem introduced in Alpern and Lidbetter \cite{alpern} and further studied in Hermans et al. \cite{hermans}. Hence, Problem A is a generalization of the expanding search problem from \cite{alpern, hermans} as well.

For any integer numbers $k, l$, $k\leq l$, let $[k:l]=\{k, k+1,...,l\}$. For each edge $e\in E$, let $c_e> 0$ be the length of the edge. An edge with the endpoints $u,v \in V$ will be denoted $(u,v)$. We use the standard scheduling notation from \cite{brucker}; so, for example, $1|\mbox{prec}|\sum w_jC_j$ denotes the non-preemptive single-machine scheduling problem of minimizing the total weighted completion time with AND-precedence constraints (AND-precedence constraints stipulate that a job can be performed only after all its predecessors are completed).

\section{Problem A on trees}

In this section we assume that network $G$ is a tree $T$; thus, $m=n-1$.  For any $u,v\in V$, let $P(u,v)$ denote the only path in $T$ with endpoints $u,v$.

\subsection{NP-hardness on a star}\label{sectionNPhard}

A \emph{tip vertex} is a vertex with degree 1. A \emph{star} is a network where one vertex (center) is incident to all edges. In a star, all vertices other than the center are tip  vertices.

We will use the following OPTIMAL LINEAR ARRANGEMENT problem which is known to be NP-complete \cite{garey, garey2}.

\emph{OPTIMAL LINEAR ARRANGEMENT}. Given an undirected graph $\tilde G=(\tilde V, \tilde E)$ with the set of vertices $\tilde V$ and a positive integer $K$, does there exist a one-to-one function $f:\tilde V\rightarrow [1:|\tilde V|]$ such that $\sum_{(u,v)\in \tilde E}|f(u)-f(v)|\le K$?

\begin{thm}\label{thmNPhard}
Problem A is strongly NP-hard even on a star.
\end{thm}

\textbf{Proof}. The proof is very similar to the proof of NP-hardness of scheduling problem $1|\mbox{prec}||\sum w_iC_i$ in Lenstra and Rinnooy Kan (\cite{lenstra}, the proof of Theorem 1). We use a reduction from OPTIMAL LINEAR ARRANGEMENT. Given an instance $(\tilde G=(\tilde V, \tilde E), K)$ of OPTIMAL LINEAR ARRANGEMENT, we construct the corresponding instance of Problem A as follows. The network will be a star with the center $o$ and $|\tilde V|$ unit-length edges $(o,v)$, $v\in \tilde V$. Let $d_v$ be the degree of vertex $v$ in $\tilde G$. The r-pairs will be $\{o, v\}$, $v\in \tilde V$, and $\{u,v\}$, $(u,v)\in \tilde E$. The weights of the r-pairs are defined as follows: $w_{\{o,v\}}=|\tilde V|-d_v$, $v\in \tilde V$; $w_{\{v,u\}}=2$, $(v,u)\in \tilde E$. It is straightforward to verify that the constructed instance of Problem A has optimal objective value not greater than $0.5|\tilde V|^2(|\tilde V|+1) + K$ if and only if the original instance of OPTIMAL LINEAR ARRANGEMENT is a ``yes"-instance. (The term $0.5|\tilde V|^2(|\tilde V|+1)$ is due to the terms $|\tilde V|$ in the weights $w_{\{o,v\}}$, which were introduced to keep the weights $w_{\{o,v\}}$ positive.)  Since all numbers involved are polynomial in the input size, this proves the theorem. $\Box$

\subsection{A dynamic programming algorithm}

In this section, we present a polynomial algorithm for Problem A on a tree with a fixed number of leaves (tip vertices).
For any subtree $T'$, let $E(T')$ be the set of edges of $T'$, let ${\cal R}(T')$ be the set of r-pairs that involve only the vertices of $T'$, let $W(T')$ be the total weight of all r-pairs from ${\cal R}(T')$, and let Problem A($T'$) be Problem A defined on the subtree $T'$ with the set of r-pairs ${\cal R}(T')$. Hence, Problem A will also be called Problem A($T$) in this section. Let $l$ be the number of leaves of $T$, and let $\Theta$ be the set of all subtrees of $T$; observe that $|\Theta|=O(n^{l})$. We assume $l$ to be a fixed constant as we are interested in developing polynomial algorithms under this assumption.

For an $e\in E$, Problem A($T$) with the additional restriction that edge $e$ is constructed last will be called Problem A$^{(e)}(T)$. For easy reference, we state explicitly the following trivial observation.

\begin{obs}\label{obsT1}
To solve Problem A($T$), it is sufficient to solve Problems A$^{(e)}(T)$ for all $e\in E$, and choose the best of the obtained $n-1$ s-sequences.
\end{obs}

The following lemma is the key to our dynamic programming algorithm.

\begin{lem}\label{lemT1} Consider an edge $e\in E$, and let $T'$ and $T''$ be the two subtrees of $T$ obtained by deleting edge $e$. Let $S(T')$ and $S(T'')$ be some optimal s-sequences for Problems A($T'$) and A($T''$), respectively. Then, there is an optimal s-sequence for Problem A$^{(e)}(T)$ where  the orders of constructing the edges of $T'$ and $T''$ are consistent with $S(T')$ and $S(T'')$. 
\end{lem}

\textbf{Proof}. See Appendix. $\Box$

Lemma \ref{lemT1} indicates that an optimal s-sequence for Problem A$^{(e)}(T)$ can be obtained by merging optimally the optimal s-sequences for smaller Problems A($T'$) and A($T''$). Such a merging procedure will be called \textbf{Procedure MERGE($e, T$)} (we will discuss later its implementation). This gives rise to a dynamic programming scheme that, starting from the trivial optimal solutions for Problem A($\cdot$) on single-edge subtrees, obtains optimal solutions for the problem on larger and larger subtrees, until an optimal solution for the whole tree $T$ is obtained. The general structure of the algorithm can be described as follows.

\begin{alg}\label{algT1}~\\
For $p=1$ to $p=n-1$\\
$~~~~~~$For all subtrees $\hat T$ of $T$ with $p$ edges\\
$~~~~~~~~~~~~$For all edges $e$ of $\hat T$\\
$~~~~~~~~~~~~~~~~~~$Solve Problem A$^{(e)}(\hat T)$ by performing Procedure MERGE($e, \hat T$);\\
$~~~~~~~~~~~~$End For;\\
$~~~~~~~~~~~~$Choose the best of the obtained s-sequences for Problems A$^{(e)}(\hat T)$, $e\in E(\hat T)$,\\ 
$~~~~~~~~~~~~$as an optimal s-sequence for Problem A($\hat T$);\\
$~~~~~~$End For;\\
End For;\\
Output the obtained optimal s-sequence for Problem A($T$).
\end{alg}

Note that Algorithm \ref{algT1} finds an optimal s-sequence for Problem A($\hat T$) for each subtree $\hat T$ of $T$; this optimal s-sequence will be denoted $S(\hat T)$.
In the remainder of this section, we will prove the following result.

\begin{thm}\label{thmT1}
Algorithm \ref{algT1}  can be implemented in $O(n^{l+2})$ time. 
\end{thm}

 First, two auxiliary observations.

\begin{obs}\label{obsT2}
All values $W(T')$ for all subtrees $T'\in \Theta$  can be computed in $O(n^l)$ time.
\end{obs}
\textbf{Proof}. See Appendix. $\Box$

For any $T'\in \Theta$ and $e\in E(T')$, let $W'(T',e)$ be the total weight of the r-pairs $\{u,v\}\in {\cal R}(T')$ such that $e\in P(u,v)$.

\begin{obs}\label{obsT21}
All values $W'(T',e)$, $T'\in \Theta$, $e\in E(T')$ can be computed in $O(n^{l+1})$ time.
\end{obs}
\textbf{Proof}. Consider $T'\in \Theta$ and $e\in E(T')$, and let $T'_1$, $T'_2$ be the subtrees of $T$ obtained by deleting edge $e$ from $T'$. Then, $W'(T',e)=W(T')-W(T'_1)-W(T'_2)$. Hence, if all values $W(T')$, $T'\in \Theta$ are known, any value $W'(T',e)$ can be computed in constant time. Taking into account $|\Theta|=O(n^l)$ and Observation \ref{obsT2} completes the proof. $\Box$

We note that Observations \ref{obsT2} and \ref{obsT21} are stronger than we need for the purposes of proving Theorem \ref{thmT1}; it would be sufficient to obtain all these values  with complexity $O(n^{l+2})$.

For any $\tilde T\in \Theta$ and $\tilde e \in E(\tilde T)$, let $W(S(\tilde T),\tilde e)$ be the total weight of the r-pairs from ${\cal R}(\tilde T)$ that become connected when edge $\tilde e$ is constructed following the s-sequence $S(\tilde T)$. Note that if $\tilde e$ is the last edge in the s-sequence $S(\tilde T)$, then $W(S(\tilde T),\tilde e)=W'(\tilde T, \tilde e)$.

Let us discuss Procedure MERGE($e, \hat T$), $e\in E(\hat T)$, for some $\hat T\in \Theta$. Let subtrees $T', T''$ be the subtrees of $\hat T$ obtained by deleting edge $e$, and suppose that $S(T')$ and $S(T'')$ have already been obtained in the course of Algorithm \ref{algT1}. According to Lemma \ref{lemT1}, imposing the constraints that the edges of $T'$ and $T''$ are constructed in the order defined by $S(T')$ and $S(T'')$, respectively, does not affect the optimal objective value of Problem A$^{(e)}(\hat T)$.

Consider the scheduling problem $1|\mbox{prec}|\sum w_jC_j$ with $|E(T')|+|E(T'')|$ jobs $J_{e'}$, $e'\in E(T')\cup E(T'')$, where job $J_{e'}$ has processing time $c_{e'}$ and weight 
\[
  w_{e'}=
     \left\{
         \begin{array}{ll}
            W(S(T'), e') & \mbox{if $e'\in E(T')$},\\
            W(S(T''),e') & \mbox{if $e'\in E(T'')$},
          \end{array}  
      \right.
\] 
with precedence constraints defined by the requirement that the order of performing the jobs must be consistent with $S(T')$ and $S(T'')$. Hence, the precedence constraints form two parallel chains. This scheduling problem will be referred to as \emph{Problem MERGE-SCHED($e,\hat T$)}. Clearly, an optimal solution (an optimal sequence of jobs and the optimal objective value) for Problem MERGE-SCHED($e,\hat T$) produces an optimal s-sequence (after appending edge $e$ at the end) and the optimal objective value (after adding $W'(\hat T, e)$) for Problem A$^{(e)}(\hat T)$. So, by performing Procedure MERGE($e, \hat T$) in Algorithm \ref{algT1} we will understand formulating and solving Problem MERGE-SCHED($e,\hat T$). We need to evaluate the effort required for formulating these problems (i.e., obtaining the necessary weights $w_{e'}$) and solving them in the course of Algorithm \ref{algT1}.

Let us now evaluate the complexity of computing all values $w_{e'}$ needed to formulate Problems MERGE-SCHED($e, \hat T$) in Algorithm \ref{algT1}. Suppose that all values specified in Observations \ref{obsT2} and \ref{obsT21} have already been computed. When $|E(T'|=1$, obtaining the value $W(S(T'),e)$ for $e\in E(T')$ is trivial. Suppose that values $W(S(T'),e)$, $e\in E(T')$ are known for all $T'\in \Theta$ such that $|E(T')|\le p-1$, and consider $\hat T\in \Theta$ such that $|E(\hat T)|=p$. Suppose that edge $\hat e$ is the last edge constructed in the s-sequence $S(\hat T)$, and let $\hat T_1$ and $\hat T_2$ be the subtrees of $\hat T$ obtained by deleting edge $\hat e$ from $\hat T$. Then, for any $e\in E(\hat T_i)$, $i=1,2$, we have $W(S(\hat T),e)=W(S(\hat T_i), e)$ which is already known; for $e=\hat e$, $W(S(\hat T),e)=W'(\hat T,e)$ which is also already known. Hence, obtaining all values required for formulating Problems MERGE-SCHED($e, \hat T$) in the course of Algorithm \ref{algT1} requires a constant time per value, and hence can be done in $O(n^{l+2})$ time (there are $O(n^{l+1})$ such problems each having $O(n)$ values).

Consider now solving Problem MERGE-SCHED($e, \hat T$) where all values $w_{e'}$ have already been obtained. This is an $1|\mbox{two chains}|\sum w_jC_j$ scheduling problem with precedence constraints in the form of two parallel chains $S_1=S(T')$ and $S_2=S(T'')$. Since chain precedence constraints are a special case of series-parallel constraints, Problem MERGE-SCHED($e, \hat T$) can be solved by the $O(n\log n)$ algorithm for Problem $1|\mbox{sp-graph}|\sum w_jC_j$ from Brucker \cite{brucker}, 5th Edition, Section 4.3.2, which would result on $O(n^{l+2}\log n)$ complexity for Algorithm \ref{algT1}. Now we show that the complexity can be improved to $O(n^{l+2})$.

Problem $1|\mbox{two chains}|\sum w_jC_j$ can also be solved by the algorithmic scheme discussed in Pinedo (\cite{pinedo}, Section 3.1) and Sidney \cite{sidney}. A group of consecutive jobs in a chain is called a \emph{block}. If a block includes the first job of the chain, it is called an \emph{initial block} of the chain. The ratio of the total weight of the jobs in a block to their total processing time is called the \emph{density} of the block. The \emph{$\rho$-factor} of a chain is the density of its maximum-density initial block. The $\rho$-factor of an empty chain is 0. 
The algorithmic scheme of Pinedo \cite{pinedo} and Sidney \cite{sidney} can be described as follows:

\begin{pro}\label{proT1}~\\
$\hat S_1:=S_1,~~\hat S_2:=S_2$;\\
Until both $\hat S_1$ and $\hat S_2$ are empty, Do\\
Begin\\
$~~~~~~$From $\{\hat S_1, \hat S_2\}$, select the chain with the highest $\rho$-factor, and perform the jobs of the initial block of this chain that defines the $\rho$-factor (which is a maximum-density initial block of the chain);\\
$~~~~~~$Delete these processed jobs from the chain;\\
End.
\end{pro}

Pinedo \cite{pinedo} does not provide a complexity analysis, and Sidney \cite{sidney} provides a complexity bound $O(n^2)$ for a somewhat more general case of $k$ chains. Using computational geometry arguments, we will prove the following result that is of some interest on its own.

\begin{lem}\label{twochains}
Problem $1|\mbox{two chains}|\sum w_jC_j$ can be solved in $O(n)$ time.
\end{lem}
\textbf{Proof}.
A \emph{density decomposition} $D(S)$ of a chain $S$ is a partition of the chain into consecutive non-overlapping blocks $B_1, B_2,...,B_k$ such that for any $i\in [1:k]$, $B_i$ is a maximum-density initial block of the chain obtained by deleting the jobs from $B_1,...,B_{i-1}$ from $S$. It is clear that after obtaining density decompositions of the chains $S_1$ and $S_2$, Procedure \ref{proT1} can be implemented in $O(n)$ time. Hence, the complexity of obtaining a density decomposition of a chain determines the complexity of problem $1|\mbox{two chains}|\sum w_jC_j$.

Consider a  chain $S=(J_1, J_2,...,J_{n'})$ with $n'$ jobs where job $J_i$ has weight $w_i$ and processing time $p_i$, $i\in [1:n']$. In the two-dimensional Euclidean plane, consider $n'+1$ points $(x_i, y_i),~i\in [0:n']$ with coordinates
\[
  x_i=\sum_{j=1}^i p_j, ~~y_i=\sum_{j=1}^i w_j,~~i\in [0:n']
\]
(observe that $x_0=y_0=0$). Since the points $(x_i, y_i)$, $i\in [0:n']$ are sorted by their  $x$-coordinates, their convex hull can be computed in $O(n')$ time by the incremental algorithm discussed in de Berg et al. \cite{berg}, Section 1.1.  Observe that the vertices of the upper boundary of the convex hull define a density decomposition of the chain $S$. Hence, a density decomposition of a chain with $n'$ jobs can be obtained in $O(n')$ time, which completes the proof of the lemma. $\Box$

Lemma \ref{twochains} with the preceding discussion imply that Algorithm \ref{algT1} can be implemented in $O(n^{l+2})$ time.  Theorem \ref{thmT1} is proven.

\begin{rem}\label{densitydecomp}
In fact, Theorem \ref{thmT1} can be proven without Lemma \ref{twochains}. In the course of Algorithm \ref{algT1}, some computations are repetitive; in particular, Procedures MERGE($e, \hat T$) at different loops may involve merging the same s-sequence $S(T')$ with different s-sequences $S(T'')$, and the density decomposition of $S(T')$ obtained in one such loop can be used in other loops where it is needed. In total, since there are $O(n^l)$ subtrees of $T$, density decomposition will have to be computed for $O(n^l)$ different chains. Hence, even if a straightforward $O(n'^2)$ algorithm is used for obtaining a density decomposition of a chain with $n'$ jobs instead of the $O(n')$ convex-hull-based algorithm discussed above, the time needed for these computations in the course of Algorithm \ref{algT1} will be only $O(n^{l+2})$. All other arguments and estimates remain the same.
\end{rem}

\begin{cor}\label{corT2}
On a path, Problem A can be solved in $O(n^4)$ time.
\end{cor}

\section{The case of a general network with a fixed number of relevant pairs}

 The results of this section are applicable not only for the total weighted connection time objective  considered so far, but also for any objective which is a non-decreasing function of the connection times of the relevant pairs. In this section, we assume that $G$ is a general network, the objective of Problem A that is being minimized is a non-decreasing function of the connection times $t_{\{u,v\}}$ of the relevant pairs $\{u,v\}\in {\cal R}$ (not necessarily the total weighted connection time considered so far), and that the objective value can be computed in $O(|{\cal R}|)$ time given the connection times of all relevant pairs. The total weighted connection time objective $\sum_{\{u,v\}\in {\cal R}} w_{\{u,v\}}t_{\{u,v\}}$ is a special case; so is the maximum lateness objective considered in \cite{averbakhJOC}. We assume that the number of relevant pairs $r=|{\cal R}|$ is a fixed constant, $r\ge 2$.

A \emph{forest} is a cycle-free network  (not necessarily connected).  A subnetwork  $\cal F$ of $G$ is an \emph{r-forest} if the following properties hold:\\
a) $\cal F$ is a forest;\\
b) Any r-pair $\{v,u\}$ is connected in $\cal F$, i.e., there is a (unique) path $P_{\cal F}(v,u)$ in $\cal F$ with endpoints $v,u$;\\
c) For any edge $e$ of $\cal F$, there is an r-pair $\{v,u\}$ such that $e\in P_{\cal F}(v,u)$.\\
These properties imply that an r-forest $\cal F$ is the union of the (unique) paths $P_{\cal F}(v,u)$, $\{v,u\}\in {\cal R}$.

For the purposes of this section, it will be convenient to introduce new terminology. The edges that are constructed before all r-pairs $\{v,u\}\in {\cal R}$ become connected are called \emph{r-essential}. The order of constructing the remaining edges does not affect the objective value. It is clear that there is an optimal schedule where the r-essential edges form an r-forest; hence we will consider only such schedules. A sequence $(e_1,...,e_f)$ of edges  that form an r-forest will be called  an \emph{rs-sequence}. Thus, the problem can be re-defined equivalently as to choose optimally an r-forest $\cal F$ (of r-essential edges) and an order of constructing its edges, i.e., an optimal rs-sequence.

For an r-forest $\cal F$, let $F({\cal F})$ be the best objective value that can be achieved using $\cal F$ as the set of r-essential edges, and let $|{\cal F}|$ denote the number of edges in $\cal F$.

\begin{obs}\label{obsR1} Consider any r-forest $\cal F$, and suppose that in some solution that has $\cal F$ as the set of r-essential edges and the objective value $F$, the r-pairs get connected in the order $(\{v_1,u_1\},...,\{v_r,u_r\})$, i.e. $t_{\{v_1,u_1\}}\le t_{\{v_2,u_2\}}\le...\le t_{\{v_r,u_r\}}$. Consider any solution with the following structure: first the edges of the path $P_{\cal F}(v_1, u_1)$ are constructed, then the remaining edges of $P_{\cal F}(v_2, u_2)$,..., then the remaining edges of $P_{\cal F}(v_r, u_r)$. Then, this solution also has $\cal F$ as the set of r-essential edges, and is not worse than the original solution, i.e. its objective value $F'$ is not greater than $F$.
\end{obs}
\textbf{Proof}. Since the obtained solution constructs first the edges of $\cal F$ and $\cal F$ is an r-forest, its set of r-essential edges is $\cal F$. Observe that the connection times of all r-pairs for the obtained solution are not greater than for the original solution. Since the objective is a non-decreasing function of the connection times of the r-pairs, we have $F'\le F$. $\Box$

Hence, for any r-forest $\cal F$, there is a solution with objective value $F({\cal F})$ and the structure described in Observation \ref{obsR1}. Since there are $r!$ possible orders of r-pairs and $r$ is fixed,   we obtain the following statement.

\begin{obs}\label{obsR2} For any r-forest $\cal F$, the value $F({\cal F})$ and the corresponding rs-sequence (best possible given that $\cal F$ is the set of r-essential edges) can be computed in $O(r!(|{\cal F}|+r))=O(|{\cal F}|)$ time. 
\end{obs}

Hence, to solve Problem A, it is sufficient to find an optimal r-forest.
There can be an exponential number of r-forests in a general network for a given set of r-pairs.  

We assume that the matrix of inter-vertex shortest path distances for $G$ is obtained as a pre-processing using the $O(n^3)$ Floyd-Warshall algorithm \cite{ahuja}, which also obtains additional information that allows us to obtain a shortest path between any two vertices in $O(n)$ time.

Suppose that ${\cal R}=\left\{\{v_1,u_1\}, \{v_2, u_2\},...,\{v_r, u_r\}\right\}$. Any r-forest is the union of some paths $P_i$ with endpoints $v_i, u_i$, respectively, $i\in [1:r]$. A vertex that belongs to some r-pair is called a \emph{terminal vertex}. In an r-forest, all tip vertices are terminal vertices (but not necessarily vice versa). For a vertex $v$ of an r-forest $\cal F$, the degree of $v$ in $\cal F$ will be called the \emph{$\cal F$-degree} of $v$. For an r-forest $\cal F$, its non-terminal vertices of $\cal F$-degree greater than 2 will be called \emph{connection vertices}, and the vertices that are either connection vertices or terminal vertices will be called \emph{significant vertices}. An r-forest can have at most $2r$ terminal vertices and at most $2r-2$ connection vertices (the latter statement can be straightforwardly shown by induction). The path in an r-forest between two significant vertices is called a \emph{basic path}, or a \emph{b-path}, if it does not contain other significant vertices. Clearly, there is an optimal r-forest where any b-path is a shortest path in $G$ between its endpoints. There can be an exponential number of shortest paths in $G$ between any pair of vertices.

Let $\hat G$ be the \emph{metric closure} of $G$, that is, a complete network with the same vertex set $V$, where the length of any edge $(u,v)$ is equal to the shortest path distance in $G$ between the vertices $u,v$. Consider Problem A on $\hat G$ with the same set of relevant pairs and the same objective; this problem will be called Problem A($\hat G$). An r-forest in $\hat G$ will be called a \emph{$\hat G$-r-forest}. For a $\hat G$-r-forest $\hat {\cal F}$, the best objective value that can be obtained by using $\hat {\cal F}$ as the set of r-essential edges for Problem A($\hat G$) will be denoted $\hat F(\hat {\cal F})$. Let $Z^*(G)$ (respectively, $Z^*(\hat G)$) be the optimal objective value for Problem A (respectively, for Problem A($\hat G$)).

For an r-forest $\cal F$, its \emph{rpr-forest} (representation forest) $\hat {\cal F}$ is a subnetwork of $\hat G$ where the vertices are the significant vertices of $\cal F$, and there is an edge between two vertices if and only if there is a b-path connecting these vertices in $\cal F$. Since $\hat {\cal F}$ is a subnetwork of $\hat G$, the length of any edge of $\hat {\cal F}$ is the length of this edge in $\hat G$. In other words, if in an r-forest we replace each b-path with an edge of the length equal to the shortest path distance in $G$ between the endpoints of this b-path, we obtain the corresponding rpr-forest. 

\begin{lem}\label{lemR1} For any r-forest $\cal F$, there is a $\hat G$-r-forest $\hat {\cal F}$  such that $ \hat F(\hat {\cal F})\le F({\cal F})$ and $\hat {\cal F}$ has no more than $2r-2$ non-terminal vertices.
\end{lem}

\textbf{Proof}. Take the rpr-forest for $\cal F$ as $\hat{\cal F}$; it is a $\hat G$-r-forest that satisfies the stated properties. $\Box$

\begin{lem}\label{lemR2} For any $\hat G$-r-forest $\hat {\cal F}$, there is an r-forest $\cal F$ such that $F({\cal F})\le \hat F(\hat{\cal F})$. This r-forest can be obtained from $\hat {\cal F}$ in $O(|\hat {\cal F}|\cdot n)$ time.
\end{lem}

\textbf{Proof}. Consider a $\hat G$-r-forest $\hat {\cal F}$, $|\hat {\cal F}|=f$. Let $\hat S=(\hat e_1,...,\hat e_f)$ be the best sequence of constructing the edges of $\hat {\cal F}$ for Problem A($\hat G$). The required r-forest $\cal F$ is obtained by replacing consecutively the edges $\hat e_1,...,\hat e_f$ with  shortest paths between their endpoints in $G$, and removing the subpaths of these shortest paths that create cycles with the already constructed part of $\cal F$. More specifically, let $SP_i$ be a shortest path in $G$ between the endpoints of $\hat e_i$, $i\in [1:f]$. The required r-forest $\cal F$ is obtained in $f$ steps starting from an empty set of edges. Let $\bar {\cal F}_i$ be the part of $\cal F$ obtained by the end of Step $i$, $i\in [1:f]$, and define $\bar {\cal F}_0=\emptyset$. In Step $i$, $i\in [1:f]$, the edges of $SP_i\setminus\bar {\cal F}_{i-1}$ are added to $\bar {\cal F}_{i-1}$; if this creates cycles, then the edges of these cycles that belong to $SP_i\setminus\bar {\cal F}_{i-1}$ are removed.  
 Each shortest path can be obtained in $O(n)$ time using the results of the pre-processing. It is straightforward to verify that the obtained set $\bar {\cal F}_f={\cal F}$ is an r-forest that satisfies the required properties. $\Box$

Lemmas \ref{lemR1} and \ref{lemR2} imply the following result.

\begin{cor}\label{corR1} $Z^*(G)=Z^*(\hat G)$, and there is an optimal $\hat G$-r-forest with no more than $2r-2$ non-terminal vertices.
\end{cor}

Observation \ref{obsR2} implies that for any $\hat G$-r-forest $\hat {\cal F}$ with no more than $2r-2$ non-terminal vertices, value $\hat F(\hat {\cal F})$ is computed in constant time (since $|\hat {\cal F}|\le 4r-3$ taking into account that there are no more than $2r$ terminal vertices). There are $O(n^{2r-2})$ $\hat G$-r-forests that have no more than $2r-2$ non-terminal vertices (since $r$ is fixed). Taking into account Corollary \ref{corR1}, we have that an optimal $\hat G$-r-forest $\hat {\cal F}^*$ can be found in $O(n^{2r-2})$ time with $O(n^3)$ pre-processing (by computing $\hat F(\hat {\cal F})$ for all $O(n^{2r-2})$ $\hat G$-r-forests with no more than $2r-2$ non-terminal vertices), and $\hat {\cal F}^*$ will have no more than $2r-2$ non-terminal vertices and therefore no more than $4r-3$ edges. Then using Lemma \ref{lemR2}, we get an optimal r-forest and an optimal rs-sequence in $O(n)$ time. 

An important special case is where all r-pairs have a common vertex; as mentioned in Section \ref{problem}, for the total weighted connection time objective this special case is equivalent to the FNCP from \cite{averbakhpereira} or the expanding search problem from \cite{alpern, hermans}. In this case, there are no more than $r-1$ connection vertices in an r-forest (which in fact will be a tree); hence, the algorithm outlined above will have complexity $O(n^{r-1})$. We obtain the following result.

\begin{thm}\label{thmR1}
On a general network, if $r$ is fixed, Problem A can be solved in $O(n^{2r-2})$ time with $O(n^3)$ pre-processing. If all r-pairs have a common vertex as in the FNCP from \cite{averbakhpereira}, then Problem A can be solved in $O(n^{r-1})$ time with an $O(n^3)$ pre-processing.
\end{thm}
\begin{rem}\label{remR1}
If  $r=2$ and both r-pairs have a common vertex, the complexity of pre-processing can be improved as in this case we do not need to obtain the full inter-vertex shortest path distance matrix for $G$. We need only the shortest path distances between the terminal vertices and all other vertices, which can be obtained by the standard Dijkstra's algorithm in $O(n^2)$ time \cite{ahuja}, or in $O(n\log n+m)$ time by Fibonacci heap implementation of the Dijkstra's algorithm \cite{ahuja}. 
\end{rem}

\section{Conclusions}
We introduced a new network construction problem (Problem A) which is a generalization of the FNCP studied in \cite{averbakhpereira, alpern, hermans}. In contrast with the FNCP which is polynomially solvable on trees, Problem A is strongly NP-hard even on stars. We presented polynomial algorithms for the cases of a tree network with a fixed number of leaves, and of a general network with a fixed number of relevant pairs.

Our focus in this paper has been on computational complexity issues. A direction for further research is to develop practical algorithmic approaches for the problem in the general case. We note that the computational approaches developed for the FNCP in \cite{averbakhpereira, hermans, averbakhCOR} do not seem  applicable to Problem A.

\vspace{0.3in}

\textbf{ACKNOWLEDGEMENT}. This research was supported by the Discovery Grant RGPIN-2018-05066 from the Natural Sciences and Engineering Research Council of Canada (NSERC). The author thanks Dr. Tianyu Wang for pointing out the reference \cite{lenstra} in connection with the proof of Theorem \ref{thmNPhard}.

\vspace{0.5in}

\textbf{APPENDIX}

\textbf{Proof of Lemma \ref{lemT1}.} Since any r-pair $\{u,v\}$ becomes connected as soon as all edges of the only path $P(u,v)$ are constructed, Problem A($T$) is equivalent to the  scheduling problem $1|\mbox{prec}|\sum w_jC_j$  with the following $|E|+|{\cal R}|$ jobs: \emph{edge-jobs} $J_e$, $e\in E$  and \emph{pair-jobs} $J_{\{u,v\}}$, $\{u,v\}\in {\cal R}$, where each edge-job $J_e$ has zero weight and processing time $c_e$ and no predecessors, and each pair-job $J_{\{u,v\}}$ has weight $w_{\{u,v\}}$ and processing time 0 and predecessors $J_e,~e\in P(u,v)$.  This scheduling problem will be referred to as \emph{Problem A-SCHED($T$)}. 

The directed acyclic graph of precedence constraints of Problem A-SCHED($T$) is bipartite (one part is the edge-jobs and the other part is the pair-jobs). Since pair-jobs have zero processing times, we can consider only solutions where any pair-job $J_{\{u,v\}}$ is performed as soon as all its predecessors $J_e, ~e\in P(u,v)$ are completed and there are no idle times, so the order of processing the edge-jobs defines a solution (the order of processing consecutive pair-jobs is immaterial since they all take zero processing time and therefore have the same completion time). Hence, a solution for Problem A-SCHED($T$) can be viewed as a sequence of edge-jobs, and there is a natural objective-value preserving correspondence between s-sequences for Problem A($T$) and solutions for Problem A-SCHED($T$). So, s-sequences for Problem A($T$) can be viewed as solutions for Problem A-SCHED($T$), and vice versa.

Consider \emph{Problem A-SCHED$^{(e)}(T)$} which is Problem A-SCHED($T$) with the additional requirement  that edge-job $J_e$ is the last performed edge-job. Problem A-SCHED$^{(e)}(T)$ is equivalent to Problem A$^{(e)}(T)$.
Suppose that  $T',~T''$ are as defined in the statement of the lemma, and an s-sequence $S$ is optimal for Problem A$^{(e)}(T)$. Then  the corresponding solution $\hat S$ for Problem A-SCHED($T$) is optimal for Problem A-SCHED$^{(e)}(T)$, and the edge-job $J_e$ and all pair-jobs $J_{\{u,v\}}$ such that $e\in P(u,v)$ are performed last in the schedule that corresponds to $\hat S$, with the same completion time $\sum_{e'\in E}c_{e'}$ which does not depend on $S$. So, the contribution of these jobs to the objective value is the same for all solutions to Problem A-SCHED$^{(e)}(T)$, hence these jobs can be ignored. After deleting these jobs, the precedence constraints  decompose into two independent parts that correspond to $T'$ and $T''$.

Let $S(T')$ and $S(T'')$ be some optimal s-sequences for Problems A($T'$) and A($T''$), respectively, and $\hat S(T')$ and $\hat S(T'')$ be the corresponding optimal solutions (edge-jobs sequences) for Problems A-SCHED($T'$) and A-SCHED($T''$). Then, Theorem 21 of Sidney \cite{sidney} implies that there is an optimal solution for Problem A-SCHED$^{(e)}(T)$ where  the order of performing edge-jobs is consistent with $\hat S(T')$ and $\hat S(T'')$. Considering the corresponding s-sequence for Problem A$^{(e)}(T)$, we have the statement of the lemma. $\Box$\\

\textbf{Proof of Observation \ref{obsT2}}. For a vertex $v\in V$ and a subtree $T'\in \Theta$, let $W_v(T')$ be the total weight of all r-pairs that contain $v$ and some vertex of $T'$. First, observe that for a fixed $v$, values $W_v(T')$ for all subtrees $T'\in \Theta$ for which $v$ is a leaf can be computed in $O(n^{l-1})$ time since each such subtree has no more than $l-1$ leaves different from $v$, and we need to spend only a constant time per subtree if we move between \emph{neighboring} subtrees that differ by only a single edge. So, values $W_v(T')$ for all $v\in V$ and all subtrees $T'\in \Theta$ for which $v$ is a leaf can be computed in $O(n^l)$ time. 

Having this information, we can compute values $W(T')$ for all subtrees $T'\in\Theta$ spending a constant time per subtree and moving between neighboring subtrees, since the difference of these values for some neighboring subtrees is value $W_v(T')$ for some $v$ and $T'$ where $v$ is a leaf of $T'$. Since $|\Theta|=O(n^l)$, this completes the proof. $\Box$

\end{document}